# Electronic properties of TaAs$_2$ topological semimetal investigated by transport and ARPES

Ashutosh S Wadge[1]*, Grzegorz Grabecki[2], Carmine Autieri[1]*, Bogdan J Kowalski[2], Przemysław Iwanowski[1,2], Giuseppe Cuono[1], M F Islam[3], C M Canali[3], Krzysztof Dybko[1,2], Andrzej Hruban[2], Andrzej Łusakowski[2], Tomasz Wojciechowski[1], Ryszard Diduszko[2], Artem Lynnyk[2], Natalia Olszowska[4], Marcin Rosmus[4], J Kołodziej[4] and Andrzej Wiśniewski[1,2]

[1]International Research Centre MagTop, Institute of Physics, Polish Academy of Sciences, Aleja Lotnikow 32/46, PL-02668 Warsaw, Poland
[2]Institute of Physics, Polish Academy of Sciences, Aleja Lotnikow 32/46, PL-02668 Warsaw, Poland
[3]Department of Physics and Electrical Engineering, Linnaeus University, 392 31 Kalmar, Sweden
[4]National Synchrotron Radiation Centre SOLARIS, Jagiellonian University, Czerwone Maki 98, PL-30392 Kraków, Poland

E-mail*: wadge@magtop.ifpan.edu.pl and autieri@magtop.ifpan.edu.pl

**Abstract**

We have performed electron transport and ARPES measurements on single crystals of transition metal dipnictide TaAs$_2$ cleaved along the ($\bar{2}$ 0 1) surface which has the lowest cleavage energy. A Fourier transform of the Shubnikov-de Haas oscillations shows four different peaks whose angular dependence was studied with respect to the angle between magnetic field and the [$\bar{2}$ 0 1] direction. The results indicate elliptical shape of the Fermi surface cross-sections. Additionally, a mobility spectrum analysis was carried out, which also reveals at least four types of carriers contributing to the conductance (two kinds of electrons and two kinds of holes). ARPES spectra were taken on freshly cleaved ($\bar{2}$ 0 1) surface and it was found that bulk states pockets at constant energy surface are elliptical, which confirms the magnetotransport angle dependent studies. First-principles calculations support the interpretation of the experimental results. The theoretical calculations better reproduce the ARPES data if the theoretical Fermi level is increased, which is due to a small n-doping of the samples. This shifts the Fermi level closer to the Dirac point, allowing investigating the physics of the Dirac and Weyl points, making this compound a platform for the investigation of the Dirac and Weyl points in three-dimensional materials.

---

wadge@magtop.ifpan.edu.pl

## 1. Introduction

After the discovery of fascinating topological insulators, the binary pnictide semimetals became also a focus of attention in the field of condensed matter. Centrosymmetric dipnictides show a large positive magnetoresistance (MR) which is one of the important reasons why these materials are intensively researched for their possible magneto-electronic applications [1]. Recently, it has been shown that the transition metal dipnictide $RX_2$ family (R = Ta, Nb; X = P, As, Sb) presents a rotational-symmetry protected topological crystalline insulator (TCI) state [2]. In particular, $TaAs_2$ shows crossings between the bands along some lines of the Brillouin zone (BZ) without spin-orbit coupling (SOC) interaction. The bands without SOC present nodal lines, while when SOC is included the nodal lines are gapped and there is band inversion at some points of the BZ. The calculation of topological invariants reveals that the system is a rotational-symmetry-protected TCI [3]. The valence and the conduction bands have different parity but both belong to d-electrons of Ta. It was shown that in the absence of magnetic field, $TaAs_2$ is a weak topological insulator, while when a magnetic field is applied type-II Weyl points have been found [4]. The presence of these Weyl points should be taken into account in the explanation of anomalous MR, quantum oscillations and negative MR experimentally observed in this compound and the other materials of this family [5].

Numerous articles report the non-saturating extremely large MR due to the presence of compensated Fermi pockets as shown by first-principle calculations and quantum oscillations [6-15], while very few cases of non-compensated materials with topological band structure show large MR. In a recent article, Butcher et al. [12] presented a detailed Fermi surface investigation of both electron and hole pockets. However, the results presented in [12] show that only electron pockets could be experimentally confirmed, whereas there is no evidence for hole pockets in the de Haas-van Alphen (dHvA) frequencies. Rao et al [16] have studied thermal conductivity and resistivity of $TaAs_2$ at ultra-low temperatures and in high magnetic fields. They found that both the resistivity and thermal conductivity display quantum oscillations at sub-kelvin temperatures and in applied fields up to 14 T. They noticed that thermal conductivity shows a $T^4$ dependence at very low temperatures, whereas the residual resistivity ratio is temperature-independent. This points to the violation of the Weidemann-Franz law and indicates a non-Fermi liquid state of this compound.

We present results of angle-resolved photo-emission spectroscopy (ARPES) studies, which according to our knowledge were not reported for this compound, as well as electron transport measurements (resistivity and Hall) performed in order to recognize the mechanism of observed extremely large MR. We also provide first-principle calculations that better reproduce the experiments if we take the theoretical Fermi level (FL) higher than the experimental one. Indeed, our samples are slightly n-doped, therefore the FL is shifted closer to the Dirac points.

The paper is organized as follows: in the next section, the details of the experimental measurements are reported; in Section III we present the computational details of our calculations. Sec. IV is devoted to results and discussion, while in Sec. V we draw our conclusions.

## 2. Experimental details

### 2.1 Single crystals growth and structural characterization

Single crystals of $TaAs_2$ were grown via the two-stage chemical vapour transport (CVT) method as shown in Fig. 1. Initially, a polycrystalline $TaAs_2$ was synthesized by a direct reaction of Ta foil (Zr industrial Ltd, 99.99%) and As (PPM Pure Metals, 99.999995%) placed inside the evacuated quartz tube at 990 ℃ for 19 days. Pellets of polycrystalline $TaAs_2$ were loaded into quartz tube with Iodine (POCH, 99.8%), sealed under the vacuum and placed into a gradient zone of temperature 1025 ℃ (crystallization zone) and 956 ℃ (source zone) for 23 days. After, the furnace was cooled down to room temperature with the rate 100 ℃/h.

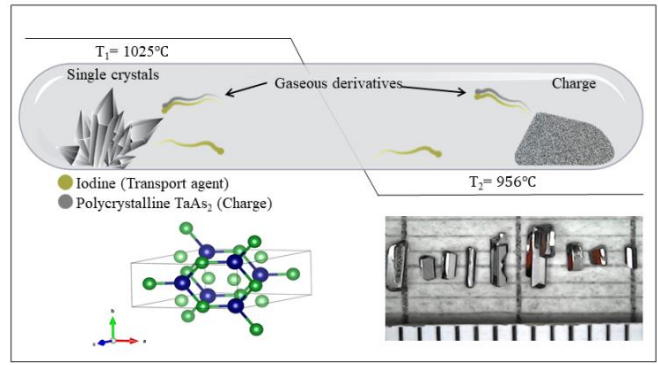

**Figure 1.** Schematic representation of Chemical Vapor transport technique used for crystallization of $TaAs_2$ in temperature gradient, insets show: the crystal structure of $TaAs_2$ (left) and millimetre sized shiny long crystals of $TaAs_2$ after the process (right).

One can describe $TaAs_2$ crystal structure by the monoclinic unit cell and C2/m (space group No. 12), which can be easily recalculated to body-centered I2/m space group. The structure of our crystals was confirmed by X-ray powder diffraction using a Rigaku SmartLab 3 kW diffractometer equipped with a tube having Cu anode, and operating with U = 40 kV and I = 30 mA. The characteristic peak positions were identified using PDXL software and data from International Centre for Diffraction Data (ICDD) as shown in Fig. 2, Powder Diffraction File PDF-4+2018 RDB database (DataBase card number: 04-003-3007). The lattice

constants determined for our crystal are a = 9.343445(12) Å, b = 3.386195(4) Å, c = 7.76195(9) Å, β = 119.7065(6)°, and V = 213,303(5) Å$^3$. The single crystal orientation was performed on a KUMA-diffraction four-circle diffractometer with a goniometer in kappa geometry. The radiation of the Cu lamp and the peak-hunting procedure were used. In a conventional unit cell of TaAs$_2$, each Tantalum atom is surrounded by six Arsenic atoms, out of those three are As1 and the remaining three are As2. Surface quality and quantitative chemical composition were verified using energy dispersive X-ray spectroscopy (EDX) system QUANTAX 400 Bruker coupled with the Zeiss Auriga field emission (Schottky type) and scanning electron microscope (FESEM), operating at 15 kV incident energy. The measurements were performed on natural planes and the results confirmed that TaAs$_2$ crystals are well stoichiometric.

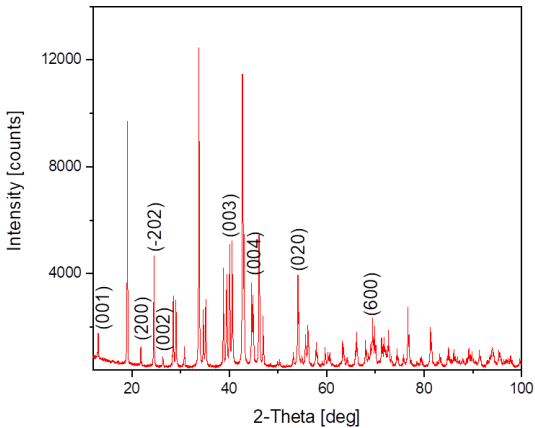

**Figure 2.** X-ray powder diffraction pattern of TaAs$_2$.

## 2.2 Electron transport measurements

Electron transport measurements were carried out in a conventional cryostat equipped with a superconducting coil capable of producing the magnetic fields up to 9 Tesla and variable temperature insert (VTI) allowing to set sample temperature in the range from 1.5 K to 310 K, with accuracy better than 0.01 K. For electrical measurements we have used AC method with acoustic frequencies with constant sample current, using lock-in technique (Stanford research SR 830) for the voltage measurements.

The TaAs$_2$ samples have been prepared in the form of rectangular Hall bars of dimensions 2 mm×0.8 mm×0.5 mm. The current was passed along the longest dimension oriented along [0 1 0] direction and the upper sample surface was perpendicular to the [$\bar{2}$ 0 1] direction along which the magnetic field was applied, see inset Fig. 3 (e). Six electric contacts were made by silver paint and gold wires of the diameter 50 μm were attached to them. We checked that the contacts were Ohmic at least up to the currents as high as several milliamperes.

## 2.3 Magnetic measurements

Magnetic properties of TaAs$_2$ sample have been investigated by means of Quantum Design Magnetic Property Measurement System MPMS XL equipped with a Superconducting Quantum Interference Device (SQUID). Reciprocating Sample Option (RSO) has been chosen to provide the precision of about 10$^{-8}$ emu during the measurements. A magnetic moment as a function of temperature has been measured in zero-field cooling ZFC and field cooling FCC modes at fixed external dc magnetic field of 10 Oe.

## 2.4 ARPES measurements

In order to investigate the Fermi surface of TaAs$_2$ single crystals experimentally, we used the UARPES beamline at the National Synchrotron Centre SOLARIS in Krakow, Poland. A variably polarizing APPLE II type undulator was the source of radiation of hν = 8 – 100 eV. The end station was equipped with the SCIENTA OMICRON DA30L photoelectron spectrometer. The energy and angular resolutions were 1.8 meV and 0.1°, respectively. It allowed for precise band-mapping in the whole Brillouin zone. ARPES spectra were taken on the ($\bar{2}$ 0 1) surface (freshly cleaved in situ in ultra-high vacuum) with excitation energy of 25 eV. The sample was kept at the temperature of 80 K. The crystallographic orientation of the cleaved surface was assessed in a separate XRD experiment. The acquired ARPES data were analyzed with the use of the software procedure package developed by the UARPES beamline staff.

## 3. Computational details

We have performed density functional theory (DFT) calculations by using the VASP package [17-19] based on the plane-wave basis set and the Projector Augmented Wave [20] method with a cutoff of 350 eV for the plane-wave basis. As an exchange-correlation functional, the generalized gradient approximation (GGA) of Perdew, Burke, and Ernzerhof (PBE) has been adopted [21]. We have used a 12×12×8 k-point Monkhorst-Pack grid for BZ integrations [2, 22] both without and with SOC. We have optimized the internal degrees of freedom by minimizing the total energy to be less than 1×10$^{-4}$ eV. After obtaining the Bloch wave functions $\psi_{n,k}$, the Wannier functions [23, 24] have been built up using the WANNIER90 code [25, 26]. Our basis for the Wannier tight-binding model is composed of the Ta-5d and As-4p. To extract the low-energy properties of the electronic bands, we have used the Slater-Koster interpolation scheme as implemented in WANNIER90. In particular, we have fitted the non-magnetic phase electronic bands to get the hopping parameters and we added the spin-orbit constants on the top to produce the Fermi surface using wanniertools [26]. In detail, the SOC values from the literature for Ta [27] and As [28] are $\lambda_{Ta}$ = 269 meV and $\lambda_{As}$ = 164 meV. Similar results were obtained also fitting the band structure including the spin-orbit coupling; these latter results are reported in this paper. The ($\bar{2}$01) surface projection of the energy spectrum and Fermi surface were obtained in a slab calculation within the iterative Green's function method implemented in Wannier tools.

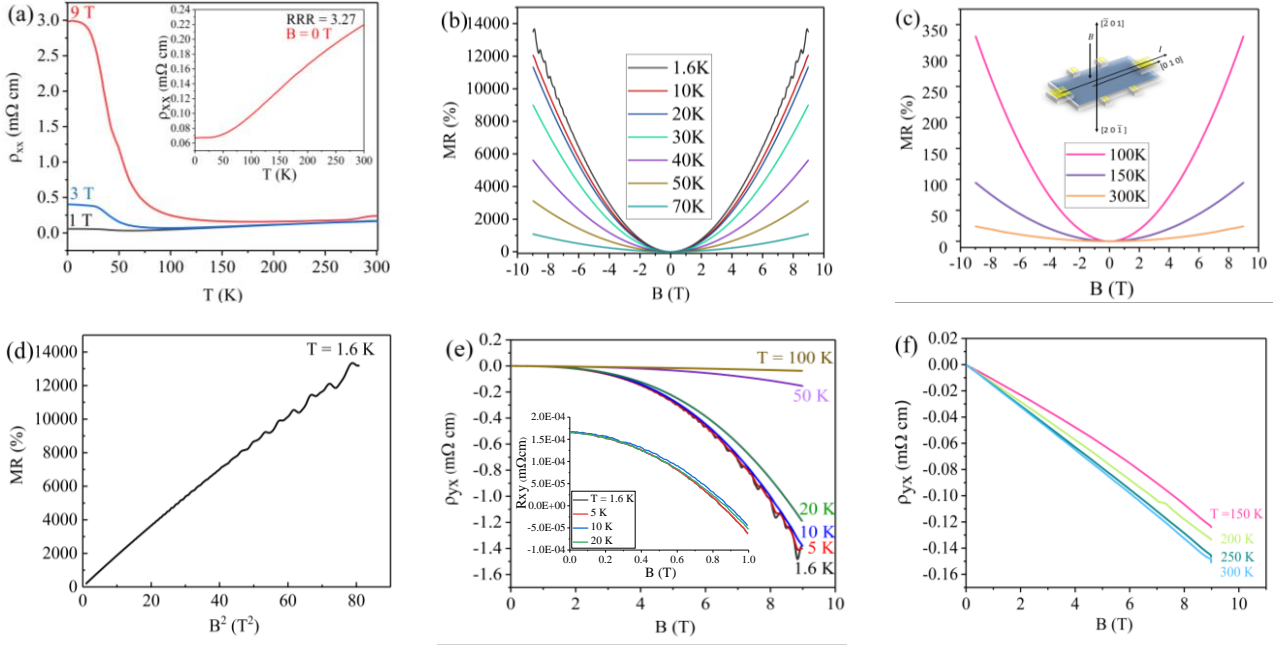

**Figure 3.** (a) Temperature dependence of resistivity at magnetic fields 1, 3 and 9 T, inset shows resistivity as a function of temperature at zero magnetic field, (b, c) MR measured by sweeping magnetic field from - 9 T to + 9 T at temperature ranges from 1.6 K to 70 K and from 100 K to 300 K, respectively, inset: schematic of sample with the direction of current in the direction of the b axis and B along [$\bar{2}$ 0 1] direction, (d) MR vs $B_2$ at 1.6 K, (e) the Hall resistivity in a magnetic field up to 9 T, in temperature ranges from 1.6 K to 100 K, inset: magnified Hall resistance showing small negative slope below 1 T for temperatures from 1.6K to 20 K (f) the Hall resistivity in magnetic field up to 9 T, in the temperature range from 150 K to 300 K, showing more linear nature.

## 4. Results and discussion

### 4.1 Electron transport - results

Temperature dependence of resistivity of the TaAs$_2$ sample, at zero magnetic field, is shown in the inset to Fig. 3(a). It shows typical metallic behaviour with residual resistivity ratio (RRR) = ρ(300 K)/ρ(1.6 K) equal to 3.65. When a perpendicular magnetic field is applied, the resistivity strongly increases with decreasing temperature, see Fig. 3(a). This is a result of the very large MR in TaAs$_2$. We have measured it in the configuration I // [010] and B // [$\bar{2}$ 0 1], and the results are shown in Fig. 3(b, c). At lower temperature, T = 1.6 K, MR is as large as $1.4 \times 10^4$ % and its magnitude shows only a minor decrease up to T = 20 K. Additionally, we observe pronounced Shubnikov-de Haas oscillations on the MR slope at the lowest temperatures, as shown in Fig. 3(d). At temperatures higher than 20 K, the MR decrease becomes faster, finally dropping down to 48% at room temperature. The Hall resistivity as a function of magnetic field was measured in the whole temperature range, from 1.6 K to 300 K. The results are presented in Fig. 3(e, f) here the consecutive curves were anti-symmetrized to remove the effect of asymmetric leads. The Hall resistance shows a strongly non-linear dependence on magnetic field, indicating the presence of several types of carriers in the sample. At lower temperatures, the Hall resistance slope is negative but very small at fields below 2 T, and strongly increases for higher fields. This clearly indicates that both the electrons and holes contribute to the total conductance. However, for temperatures larger than 100 K, the Hall resistance is almost constant in the whole field range, see Fig. 3(f). This indicates that the contribution of electrons increases with increasing temperature.

Due to the multicarrier nature of conductivity in TaAs$_2$, the determination of the mobilities and concentrations for each type of carriers requires an application of additional procedures. As a first step, we have used the two band model to determine averaged electron and hole concentrations in the sample, in a way similar to that in Ref [4]:

$$\rho_{xx} = \frac{\sigma_e + \sigma_h + \sigma_e\sigma_h(\sigma_e R_e^2 + \sigma_h R_h^2)B^2}{(\sigma_e + \sigma_h)^2 + \sigma_e^2\sigma_h^2(R_e + R_h)^2 B^2} \quad (1)$$

$$\rho_{xy} = B\left(\frac{R_e\sigma_e^2 + R_h\sigma_h^2 + \sigma_e^2\sigma_h^2 R_e R_h(R_e + R_h)B^2}{(\sigma_e + \sigma_h)^2 + \sigma_e^2\sigma_h^2(R_e + R_h)^2 B^2}\right) \quad (2)$$

The results were fitted to the experimental data measured at T = 1.6 K and are shown in Fig. 4(a). From the fit, we have determined the hole concentration $n_h = 3.1 \times 10^{19}$ cm$^{-3}$ and electron concentration $n_e = 3.5 \times 10^{19}$ cm$^{-3}$. The corresponding mobilities are $1.4 \times 10^4$ cm$^2$/Vs and $1.3 \times 10^4$ cm$^2$/Vs.

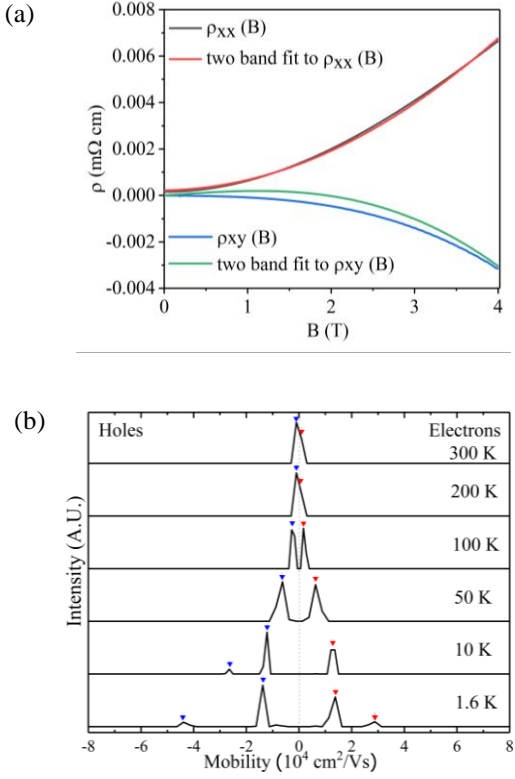

**Figure 4.** (a) Two band model fit of the Hall resistivity at 1.6 K, (b) Mobility spectrum analysis showing electron (red arrows) and hole (blue arrows) mobilities in the temperature range from 1.6 K to 300 K.

It has to be pointed out, however, that the above method gives only averaged values in the case where more than one types of electrons or holes are present in the material. Therefore, to obtain more precise data, we have performed mobility spectrum analysis (MSA) [29-32]. It is an efficient technique to characterize systems with more than two types of carriers. MSA considers the values of $\rho_{xx}(B)$ and $\rho_{xy}(B)$ components. These values are capable of handling different carrier species with respect to their average mobility which changes with change in magnetic field value. The magneto-conductivity tensor components $\sigma_{xx}(B)$ and $\sigma_{xy}(B)$ are calculated as follows:

$$\sigma_{xx}(B) = \frac{\rho_{xx}(B)}{[\rho_{xx}(B)]^2 + [\rho_{xy}(B)]^2} \quad (3)$$

$$\sigma_{xy}(B) = \frac{\rho_{xy}(B)}{[\rho_{xx}(B)]^2 + [\rho_{xy}(B)]^2} \quad (4)$$

Tensor components, as described by Eq. (3) and (4), are used to calculate conductivity density $S(\mu)$ which is given as [12]:

$$\sigma_{xx}(B) = \int_{-\infty}^{\infty} \frac{S(\mu)}{1+(B\mu)^2} d\mu \quad (5)$$

$$\sigma_{xy}(B) = \int_{-\infty}^{\infty} \frac{S(\mu)\mu B}{1+(B\mu)^2} d\mu \quad (6)$$

The mobility spectrum obtained by MSA consists of separate peaks corresponding to different types of carriers. We have applied this technique for our experimental data based on a program written in a FORTRAN platform with 800 points. We have excluded the data after 4 T in magnetoresistance and Hall resistivity because SdH oscillations have a detrimental impact on this analysis. We have observed distinct peaks for which negative values denote hole-type carriers and positive ones electron-type. The mobility spectrum at 1.6 K shows four peaks suggesting the presence of four types of carriers as shown in Fig. 4(b). They gradually merge into one as the temperature rises.

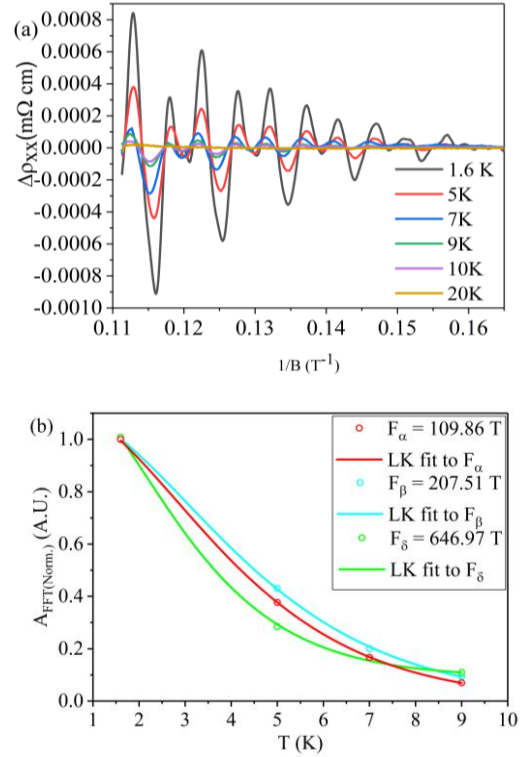

**Figure 5.** (a) Shubnikov-de Haas oscillations measured in temperature range from 1.6 K to 20 K (the monotonic background was subtracted), for T ≥ 20 K quantum oscillations vanish, (b) Lifshitz-Kosevich fit to the temperature dependence of FFT peaks magnitude.

A powerful tool for determining different carrier species in multi-carrier systems is the observation and analysis of the

**Table 1.** Transport parameters determined from SdH oscillations with B // [$\bar{2}$ 0 1] direction

| Frequency (T) | Effective mass (m*) | Area of cross-section of Fermi surface (Å$^{-2}$) | Fermi wave vector (Å$^{-1}$) | Fermi velocity (m/s) | Fermi energy (meV) |
|---|---|---|---|---|---|
| 109 | 0.30 m$_e$ | 0.01 | 0.06 | 2.2×10$^5$ | 42 |
| 207 | 0.28 m$_e$ | 0.02 | 0.08 | 3.3×10$^5$ | 87 |
| 646 | 0.40 m$_e$ | 0.06 | 0.10 | 2.9×10$^5$ | 189 |

Shubnikov-de Haas (SdH) oscillations. This is because the oscillation periods in 1/B are proportional to the area of the Fermi surface cross-sections perpendicular to the magnetic field direction. Therefore, one can directly show the presence of different carrier species and their concentrations. We observed pronounced SdH oscillations in both ρ$_{xx}$ and ρ$_{xy}$, well visible on the magnetoresistance and Hall resistance slopes, see Fig. 3(b, e). After subtracting the background signal, we plotted Δρ$_{xx}$ vs the inverse of magnetic field (B$^{-1}$), see Fig. 5(a). The oscillation amplitude strongly decreases with the temperature and at 20 K virtually vanishes. We analyzed the oscillation patterns by means of the fast Fourier transform (FFT). The temperature dependence of these oscillation components was used to calculate the cyclotron masses for different carrier species. We have plotted the thermal damping of FFT amplitude A$_{FFT}$(T), as shown in Fig. 5(b), where we normalized A$_{FFT}$(T) to the lowest amplitude A$_{FFT}$(T$_{1.6K}$). Cyclotron mass can be extracted by Lifshitz-Kosevich formula [33]:

$$\Delta A_{FFTNorm}(T) = \frac{2\pi^2 k_B m^* T / e\hbar B}{\sinh\left(2\pi^2 k_B m^* T / e\hbar B\right)} \quad (7)$$

Where: $k_B$ is Boltzmann constant, $\hbar$ is a reduced Planck constant, e is the charge of electron, mean value of 1/B interval and T is the temperature. Here, we have plotted temperature dependence of frequency components and extracted effective masses and other related parameters such as area of cross-section of Fermi surface, Fermi wave vector (assuming a circular cross-section), Fermi velocity and Fermi energy which are shown in Table 1.

(a)
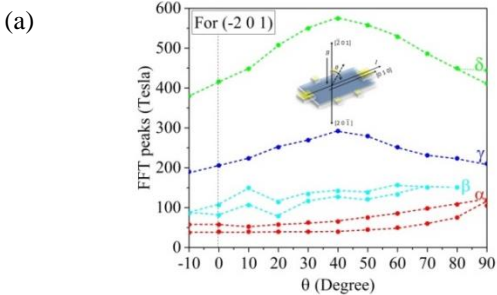

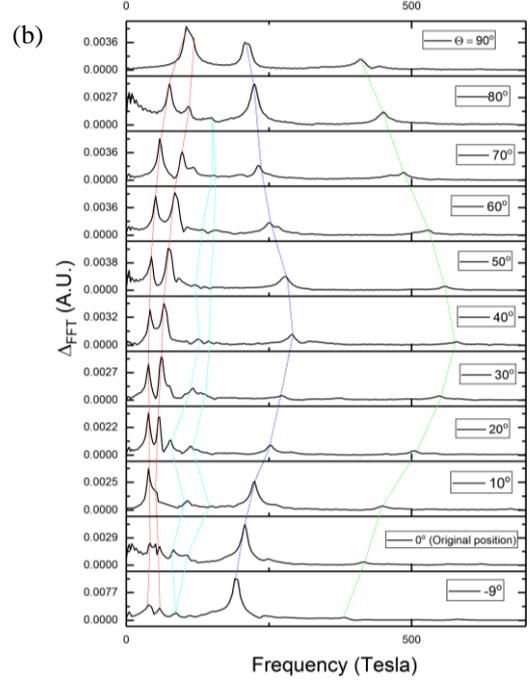

**Figure 6.** (a) Angle dependence of Shubnikov de Haas frequencies measure by rotating ($\bar{2}$0 1) plane with reference to magnetic field (b) FFT peak positions as a function of angle (θ) between B and [$\bar{2}$ 0 1].

We have studied the shapes of the cross-sections of the Fermi surfaces by changing the angle between ($\bar{2}$ 0 1) surface of the sample and magnetic field. For this purpose, we used a sample holder equipped with a rotary mechanism that allows to change the angle with an accuracy better than 1°. The configuration of magnetic field and electric current are as shown in outset Fig. 6 (a, b). We have systematically performed FFT analysis for SdH oscillations measured at each angle. The observed variance of the FFT peak positions clearly shows anisotropy of the SdH frequencies which can be produced by anisotropic Fermi surfaces.

We have performed measurements of the parallel magnetoresistance (sample geometry in parallel configuration shown in inset of Fig. 7(a)) in a temperature range from 1.6 to 10 K. We observed that at zero magnetic field there is a sharp intense peak at 1.6 K which reduces as temperature increases and finally vanishes above 10 K, as represented in Fig. 7(a).

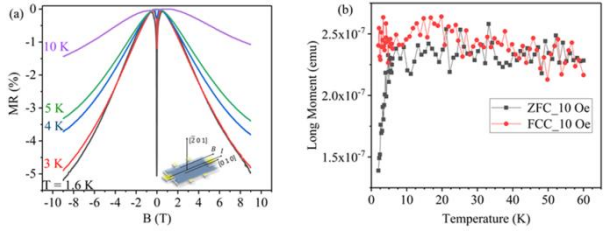

**Figure 7.** (a) Negative MR from 1.6 K to 10 K showing a sharp peak at zero magnetic field, at 10 K the peak vanishes, inset shows contact geometry of the sample in which the direction of magnetic field (B) is parallel to the direction of current (I), (b) zero field cooling (ZFC - gray color) and field cooling (FCC - red color) at 10 Oe field showing transition at about 4.5 K.

It might be due to the existence of some small superconducting clusters of Tantalum, as it was also confirmed by magnetization measurements whose results are represented in Fig. 7(b). The origin of the negative MR in $TaAs_2$ is the presence of the Zeeman effect induced Weyl points enclosed within these Fermi surfaces. This usually happens for materials that have a non-zero Chern number under magnetic field [3].

### 4.2 Electron transport - discussion

It is generally accepted that the residual resistance ratio (RRR) is a measure of the purity of metallic conductors. Our experimental data show RRR = 3.65, which is significantly smaller than those reported by others [6, 7, 10, 12]. However, studied samples show the magnetotransport properties, qualitatively the same as reported in the previous papers. In particular, the perpendicular magnetoresistance reaching $1.4 \times 10^4$% at 9 T and 1.6 K, is smaller by one order of magnitude than published in [5, 11], but still preserves its quadratic dependence on B. One of the reasons may be significantly higher carrier concentrations in our samples, which are by 2-3 times higher than those studied so far. We have $n_e = 3.5 \times 10^{19}$ cm$^{-3}$ and $n_h = 3.1 \times 10^{19}$ cm$^{-3}$, but for example Luo et al. [5] Reported $n_e = 1.4 \times 10^{19}$ cm$^{-3}$ and $n_h = 1.0 \times 10^{19}$ cm$^{-3}$. Similarly, Wu et al. [10] found $n_e = 2.790 \times 10^{18}$ cm$^{-3}$ and $n_h = 2.786 \times 10^{18}$ cm$^{-3}$. Despite these quantitative differences, we also observe well-developed quantum oscillations, which indicate that our samples are suitable for studying the band structure properties of $TaAs_2$. It has also to be noted that SdH frequencies obtained in the present work show similarity to those observed previously. For example, Butcher et al [12] observed for the same sample orientation FFT peaks at 85 and 130 T, which may be analogous to ours α and β, but for lower electron concentrations.

An additional method of analysis of the transport data applied in the present work is Mobility Spectrum Analysis. In contrast to the analysis using equations (1-2), it allows separating different carrier species with the same charge, either electrons or holes. Indeed, in our sample, we obtain well-defined mobility peaks for four types of carriers, two for electrons and two for holes. As the temperature is raised, the peak positions move towards zero and consecutively merge into one peak. This is a result of mobility decrease caused by the scattering due to phonons.

### 4.3 Band structure with experimental lattice constants

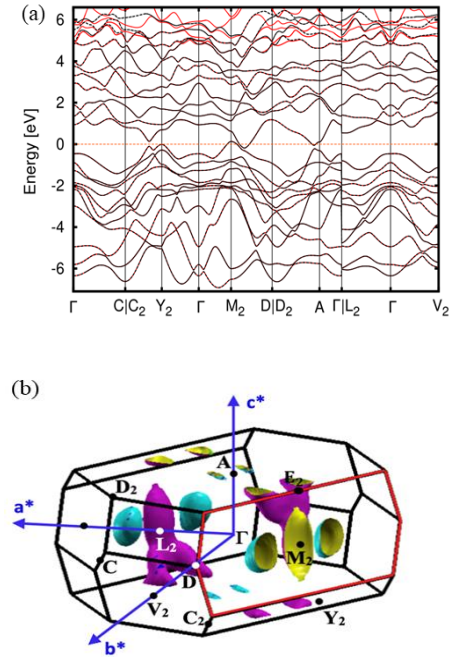

**Figure 8.** (a) Band structure of $TaAs_2$ with SOC obtained with the experimental lattice constants and plotted in the range between -7.5 eV and 6.5 eV. In red the DFT result is presented, the black dotted lines indicate the p-d tight-binding obtained with Wannier90. The Fermi level is set at zero energy. (b) Fermi surface of $TaAs_2$ with SOC obtained with the experimental lattice constants. The red lines delimit the surface related to the orientation $(\bar{2}\,0\,1)$ that is the relevant orientation for the ARPES experiments.

The experimental lattice constants of the primitive cell of the $TaAs_2$ at room temperature are $a_P = 4.963$ Å, $b_P = 4.963$ Å, $c_P = 7.752$ Å, the angles are $\alpha_P = 62.229°$, $\beta_P = 117.771°$, $\gamma_P = 140.143°$. In the convention used by the experimentalists, the relations between the lattice vectors of the primitive and the conventional unit cell are: $\vec{a_C} = \vec{a_P} - \vec{b_P}$, $\vec{b_C} = \vec{a_P} + \vec{b_P}$, $\vec{c_C} = \vec{c_P}$. As a consequence of the previous relations, the lattice constants of the conventional unit cell are $a_C = 9.331$ Å, $b_C = 3.383$ Å, $c_C = 7.752$ Å, while the angles are $\alpha_C = 90°$, $\beta_C = 119.71°$, $\gamma_C = 90°$. Since the ARPES experiments were performed at 80 K, we would need lattice constants as small as possible for an agreement with experiments. We have investigated the transition metal pnictide and we found that the Ta is $s^0d^5$. In these compounds, the oxidation states are zero when the stoichiometry of the transition metal is comparable with the pnictide like in $TaAs_2$, while the ionicity and oxidation states increase with the increase of the percentage of transition-metal [34-36]. We have chosen the k-path proposed in [37]. In Fig. 8(a), we report the bands with SOC obtained with DFT calculations and the fit with Wannier 90 in a wide energy range around the Fermi level. In Fig. 8(b) we report the three dimensional Fermi surface of the system and the Brillouin zone, with highlighted the surface related to the orientation ($\bar{2}$ 0 1) that is the relevant orientation for the ARPES experiments.

The Wannier tight-binding Hamiltonian gives a perfect match with the DFT band structure up to 5 eV above the Fermi level; above 5 eV there is hybridization with the 6s orbitals of Ta. The not-perfect agreement of the band structure above 5 eV produces a shift of the Wannier centers and makes it impossible to estimate the on-site spin-orbit coupling from the tight-binding. Adding the value of SOC to the non-magnetic TB model, we obtain an excellent agreement with the DFT SOC band structure and we use this to calculate the Fermi surface shown later in the paper.

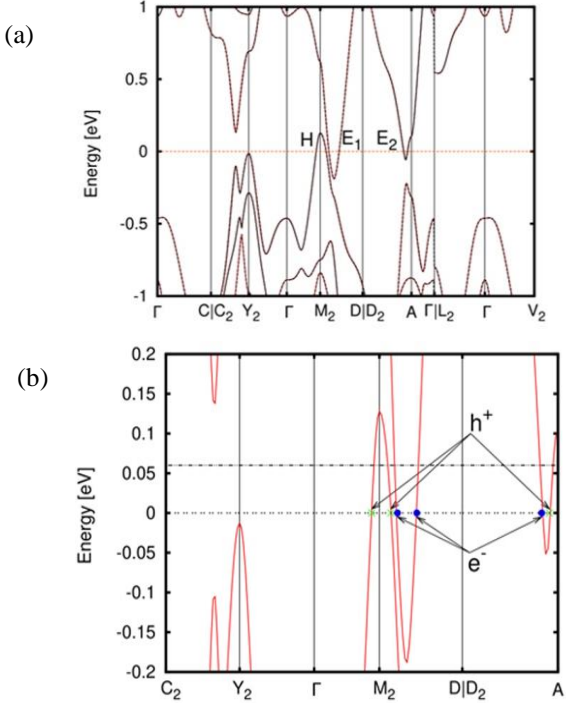

**Figure 9.** Band structure of $TaAs_2$ with SOC obtained with the experimental lattice constants. (a) In red there is the DFT result, the black dotted lines indicate the p-d tight-binding obtained with Wannier90 that matches completely the DFT band structure in the presented energy range. The plot is in the range between -1 eV and 1 eV, the Fermi level is set at zero energy. (b) Magnification of band structure with electron and hole effective masses indicated by blue and green markers, respectively.

The Fermi level is crossed by three bands: one band around the $M_2$ point, one band between the $M_2$ and D and the last one between $D_2$ and A. Using the notation of reference [12], these bands give rise to the Fermi sheets H, E1 and E2, respectively; see Fig. 9(a). The first band is hole-like; the second is electron-like while the last one has a mixed character between hole and electron due to band inversion, see Fig. 9(b). Indeed, from the minimum of the band towards $D_2$ we have an electron-like curvature, while from the minimum of the band towards A, we have a hole-like curvature.

### 4.4 Topological surface states

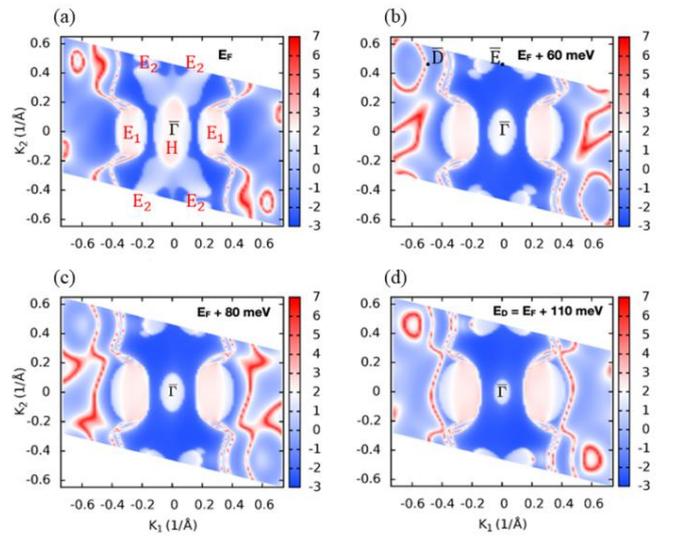

**Figure 10.** Theoretical Fermi surface of the $TaAs_2$ slab with ($\bar{2}$01) orientation as a function of the Fermi level. In panel (a) the theoretical Fermi level was used, while in panel (b) and (c) the Fermi level was increased by 60 and 80 meV, respectively. In panel (d) the Fermi surface at the energy of the Dirac points is shown. The colorbar indicates the spectral function in logarithmic scale, therefore the blue color denotes zero states, the white color denotes the bulk states while the red lines denote the surface states. The best agreement between theory and experiments is achieved for an energy shift of 60 meV above the theoretical Fermi level for the pocket at the $\bar{\Gamma}$ point. The $\bar{\Gamma}$ point is at the coordinates $K_1 = K_2 = 0$, the center of the figures. Also the points $\bar{D}$ and $\bar{E}$, the projection of the D and E points on the ($\bar{2}$01) surface, are shown. From the Brillouin zone of the primitive unit cell, the shape of the 2D Fermi surface is a non-regular hexagon, but it was plotted in an equivalent rhombus shape for better visualization.

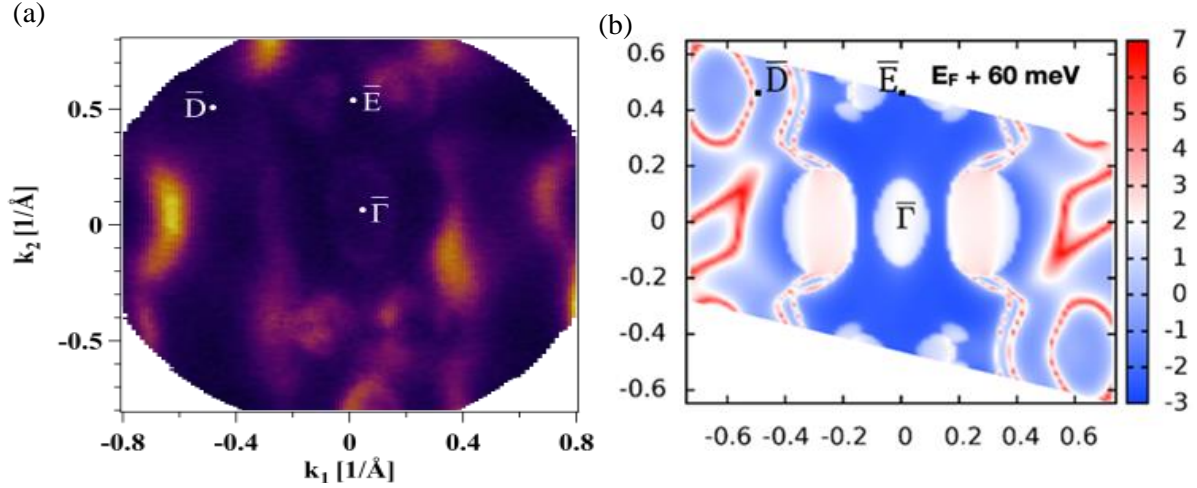

**Figure 11.** (a) Experimental constant energy contour taken at photon energy 25 eV (best case of energy resolution), (b) theoretical Fermi surface at the Fermi level raised slightly towards conduction band and theoretical Fermi level, respectively. In the theoretical figures, the colorbar indicates the spectral function on a logarithmic scale. The best agreement between theory and experiments is achieved for a shift of the Fermi level by +60 meV.

We have investigated theoretically the semi-infinite slab with orientation ($\bar{2}$ 0 1) that is the relevant orientation for the ARPES experiments. We have calculated the Fermi surface tuning the Fermi level as shown in Fig. 10. The result in Fig. 10 (a) represents the Fermi surface without doping. Since the system is not a pure insulator, the bulk bands are also projected onto the surface states, therefore, we have a mix of trivial and topological surface states. The bulk contribution to the Fermi surface is composed of 7 pockets, these 7 pockets can be described with three Fermi surfaces. Using the notation of reference [12] we have two electron pockets $E_1$ on both sides of the k-point $M_2$ in the bulk notation (or $\bar{\Gamma}$ since k-point $M_2$ is projected on $\bar{\Gamma}$ in the surface notation). We have 4 electron pockets $E_2$ on both sides of the k-point A and one hole pocket centered at the k-point $M_2$. We define $K_1$ and $K_2$ as the coordinate of the k-space of the ($\bar{2}$ 0 1) surface orientation as shown in Fig. 10. Two electron pockets $E_1$ are centered at $(K_1, K_2) \approx (\pm 0.25\ \text{Å}^{-1}, 0)$, while the four electron pockets $E_2$ are centered at $(K_1, K_2) \approx (\pm 0.15\ \text{Å}^{-1}, \pm 0.45\ \text{Å}^{-1})$. The Fermi surface $E_2$ has also some hole character between $\bar{E}$ and $\bar{\Gamma}$.

Regarding the topological surface states, the system presents the Dirac point at circa 110 meV above the Fermi level [2]. We have tuned the Fermi level scanning from the undoped Fermi level until the Dirac point. The Fermi surface with a shift of +60, +80 and +110 meV has been reported in Figs. 10 (b,c,d), respectively. With increasing the Fermi level, the hole pocket at $\bar{\Gamma}$ shrinks while the hole character of the $E_2$ pocket disappears leaving $E_2$ just with its electron character. In Figs. 11 (a) we show the experimental Fermi surface, while in Figs. 11 (b) we report the theoretical one with the shift of 60 meV of the Fermi level. The 2D Brillouin zone of Figs. 11 (b) is equivalent to the one highlighted in Fig. 8(b). As far as the authors know, this is the first ARPES for the topological semimetal TaAs$_2$.

If we increase the Fermi level with respect to the theoretical value by 60 meV, we find a better agreement with the experimental results for the hole pocket at the $\bar{\Gamma}$ point, while the pockets at $K_2 \approx \pm 0.45$ Å$^{-1}$ are better reproduced at 110 meV above Fermi level. While we have a reasonable agreement for bulk states, some surface states are also visible but not clear in the experimental figure, as shown in Figs. 11 (a), probably due to the not perfect quality of the surface of these compounds that are difficult to cleave. The shift of the Fermi level by 60 meV makes the Fermi level closer to the Dirac points. The Dirac points in zero magnetic field are present at 110 meV above the undoped Fermi level (see Fig. 11); therefore, the experimental Fermi level is in between the theoretical Fermi level and the Dirac points. This small n-doping gets closer to the experimental Fermi level to the Dirac points. This allows us to investigate the physics of Dirac point at zero magnetic field and the physics of Weyl points in applied magnetic field. From the comparison of the size of the Fermi surfaces with Table I and the theoretical Fermi pockets, we can associate the three frequencies corresponding to frequencies 109, 207 and 646 T to the E2, H and E1 Fermi surfaces, respectively.

From the DOS, we have found that the number of extra electrons per formula unit to shift the Fermi level by +60 meV is 0.020 electrons. This value is relatively small due to the low DOS of this semimetallic phase. Moreover, we have an electron concentration $n_e = 3.5 \times 10^{19}$ cm$^{-3}$ slightly larger than hole concentration $n_h = 3.1 \times 10^{19}$ cm$^{-3}$. Therefore, it is reasonable to assume that we have slightly n-doped samples. The DFT calculated carrier concentrations overestimate the experimental values due to the difficulty of the theory to catch a relatively small value of the carrier concentration.

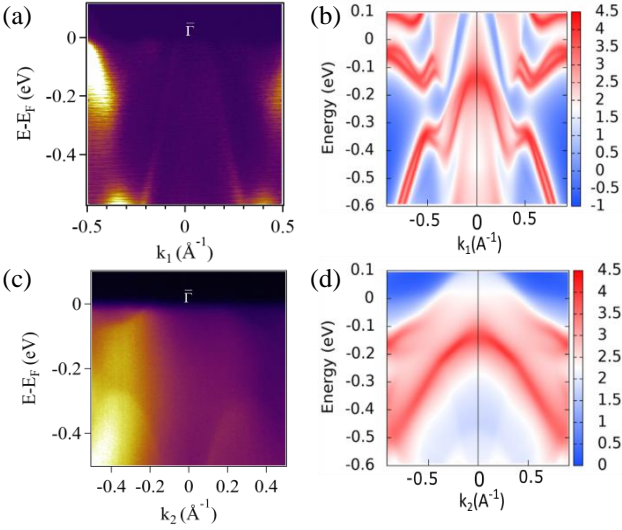

**Figure 12.** (a, c) Experimental electronic structure at $K_2 = 0$ and $K_1 = 0$, (b, d) theoretical electronic structure at $K_2 = 0$ and $K_1 = 0$, respectively.

We have plotted the experimental and theoretical electronic structure of the semi-infinite slab in Fig. 12 along two high-symmetry lines ad $K_1=0$ and $K_2=0$. Figs. 12(a) and (b) represent the experimental and theoretical electronic structure at $K_2=0$, where we observe surface states at the zone boundary between -0.2 eV and the Fermi level in both theoretical and experimental data. Other surface states at the zone boundary appear at -0.6 eV in the experimental figure, while they appear at -0.3 eV in the theoretical data. Figs. 12(c) and (d) represent the direction at $K_1=0$ where fewer features are present and we can observe just the band that generates the hole pocket H.

In general, these high-symmetry dispersions present a good agreement between theory and experiments. The experimental data clearly show the presence of surface states compatible with the surface states observed in the theoretical band structure that predict the topological character of this compound.

*4.5 Cleavage energy*

In this study, we were able to cleave a particular surface ($\bar{2}$ 0 1). The notation of the crystal direction was done in the conventional unit cell. We have calculated the cleavage energy (CE) using DFT. Using the notation of the conventional unit cell, we obtain CE$_{(\bar{2}01)}$ = 1.47 eV, CE$_{(001)}$ = 1.62 eV and CE$_{(110)}$ = 3.78 eV. The lowest value of the CE$_{(\bar{2}01)}$ confirms that the ($\bar{2}$ 0 1) surface is the easiest to cleave. The particularity of this ($\bar{2}$ 0 1) surface is the exclusive presence of As-As bonds as shown in Fig. 12(a). We associate this low CE to the presence of As-As bond, also the (001) surface with low cleavage energy shows many As-As bonds confirming this effect. We assume that the bond As-As is weaker than the Ta-As bond producing lower cleavage energy making it possible to cleave this particular surface. We attribute the physical reason to the presence of As p-states that are far from the Fermi level, therefore the As p-states are weakly bonding states compared with the Ta states that are at the Fermi level. The core-level spectra taken at $h\nu = 100$ eV show the characteristic tantalum and arsenic peaks confirming the clean cleaved surface (see Fig. 13b).

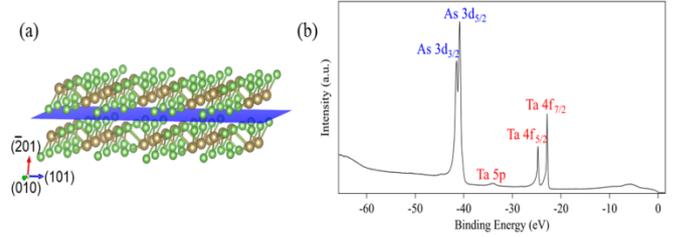

**Figure 13.** (a) Lateral view of the ($\bar{2}$01) surface. The blue plane represents the ($\bar{2}$01) plane. The axes are in the notation of the conventional unit cell. The green and brown balls represent the As and Ta atoms, respectively. (b) Core level spectra taken at hv = 100 eV showing Tantalum core level peaks at 4f and 5p (red color) as well as 3d core level peak of Arsenic (blue color).

## 5. Conclusions

We have performed electron transport and ARPES measurements on TaAs$_2$ semimetal crystals cleaved along ($\bar{2}$ 0 1) surface which has the lowest cleavage energy. In spite of values of the residual resistance ratio and the magnetoresistance smaller than the ones reported in previous works, our crystals show pronounced Shubnikov-de Haas oscillations at helium temperatures, allowing us to analyze details of the band structure of the material. In particular, the oscillations show pronounced anisotropy when the magnetic field direction is varied with respect to ($\bar{2}$ 0 1) axis, indicating non spherical Fermi surfaces, consistently with ARPES results and the ab-initio calculations. Additionally, the mobility spectrum analysis of the conductivity tensor, σ$_{xx}$(B) and σ$_{xy}$(B), shows evidence that two kinds of electrons and two kinds of holes participate in the total conductance. This indicates that in TaAs$_2$ we observe strong compensation of the conducting carriers which is also supported by ab initio calculations.

ARPES enabled us to map the electronic band structure of TaAs$_2$. The constant energy cuts parallel to the ($\bar{2}$ 0 1) plane were compared with the corresponding Fermi surface diagrams obtained by the first-principle calculations.

The main features revealed in the experimental band structure cuts can be correlated with those found in the calculated Fermi surface images, which can be interpreted both as surface states or bulk states projected onto the surface. Therefore, a mix of trivial and topological surface states is observed in the ARPES results. In the experimental diagrams, seven Fermi surface pockets can be discerned, in agreement with the bulk pockets of the theoretical Fermi surface. The bulk states pockets at the constant energy surfaces are elliptical, in agreement with the first-principles calculation results. These results are also consistent with the outcome of the transport angle-dependent studies.

The experimental and calculated Fermi surface diagrams were compared with the use of the Fermi level position as an adjustment parameter. We achieved the best agreement between the theoretical Fermi surface and the ARPES data if we compare the ARPES data taken at the Fermi energy of the sample with the theoretical Fermi surface corresponding to the energy of 60 meV above the zero-energy level of the calculated band structure. The experimental electronic structure clearly shows the presence of surface states compatible with the topological surface states observed in the theoretical band structure. Since the transport study of the investigated crystals proved that the electron concentration is slightly larger than the hole concentration, such a Fermi position, corresponding to a slightly n-type conductivity, is reasonable. The evolution of the Fermi surfaces as a function of energy is also consistent in the calculated band structure and the ARPES data. In particular, the hole pocket at $\bar{\Gamma}$ shrinks with increasing energy leaving, for energies higher than 120 meV, just the electron pockets. TaAs$_2$ exhibits a Dirac point at 110 meV. The experimental Fermi level is in between the theoretical Fermi level and the Dirac point. This allows us to investigate the physics of Dirac point at zero magnetic field and the physics of Weyl points in applied magnetic field.

## Acknowledgements

The work is partially supported by the Foundation for Polish Science through the International Research Agendas program co-financed by the European Union within the Smart Growth Operational Programme. We acknowledge the access to the computing facilities of the Interdisciplinary Center of Modeling at the University of Warsaw, Grant No. G84-0, GB84-1 and GB84-7. We acknowledge the CINECA award under the ISCRA initiatives IsC85 "TOPMOST" grant, for the availability of high-performance computing resources and support. CMC and MdF were supported by the Faculty of Technology at Linnaeus University and by the Swedish Research Council under Grant Number: 621-2014-4785. Computational resources at Linnaus University were provided by the Swedish National Infrastructure for Computing (SNIC) at Lunarc partially funded by the Swedish Research Council through grant agreement no. 2018-05973.